\documentclass[twocolumn,showpacs,nofootinbib,prd]{revtex4}
\usepackage{graphicx}
\usepackage{dcolumn}
\usepackage{bm}
\usepackage{color}

\newcommand{\One}{1\!\!1}
\newcommand{\rse}{\mathcal{R}}
\newcommand{\bes}[1]{j^{#1}_{L_{#1}}}
\newcommand{\han}[1]{h^{(1)#1}_{L_{#1}}}

\newcommand{\tmat}[2]{T_{#1#2}^{(L_{#1},L_{#2})}}
\newcommand{\tso}{$^{3\!}S_1$}
\newcommand{\stso}{$2\,{}^{3\!}S_1$}
\newcommand{\ttso}{$3\,{}^{3\!}S_1$}
\newcommand{\ftso}{$4\,{}^{3\!}S_1$}
\newcommand{\tdo}{$^{3\!}D_1$}
\newcommand{\ftdo}{$1\,{}^{3\!}D_1$}
\newcommand{\stdo}{$2\,{}^{3\!}D_1$}
\newcommand{\ttdo}{$3\,{}^{3\!}D_1$}
\begin{document}

\title{Multichannel calculation of excited vector $\phi$ resonances and the
$\phi(2170)$}

\author{Susana Coito}
\email{susana.coito@ist.utl.pt}
\affiliation{Centro de F\'isica das Interac\c c\~oes Fundamentais, Instituto Superior
T\'ecnico, Technical University of Lisbon, P-1049-001 Lisboa, Portugal
}
\author{George Rupp}
 \email{george@ist.utl.pt}
\affiliation{Centro de F\'isica das Interac\c c\~oes Fundamentais, Instituto Superior
T\'ecnico, Technical University of Lisbon, P-1049-001 Lisboa, Portugal
}
\author{Eef van Beveren}
\email{eef@teor.fis.uc.pt}
\affiliation{
Centro de F\'{\i}sica Computacional, Departamento de F\'{\i}sica,
Universidade de Coimbra, P-3004-516 Coimbra, Portugal
}

\date{\today}

\begin{abstract}
A multichannel calculation of excited $J^{PC}=1^{--}$ $\phi$ states is carried
out within a generalization of the Resonance-Spectrum Expansion, which may
shed light on the classification of the $\phi(2170)$ resonance, discovered by
BABAR and originally denoted $X(2175)$. In this framework, a complete spectrum
of bare $s\bar{s}$ states is coupled to those OZI-allowed decay channels that
should be most relevant for the considered energy range. The included $S$- and
$P$-wave two-meson channels comprise the lowest pseudoscalar, vector, scalar,
and axial-vector mesons, while in the $q\bar{q}$ sector both the $^{3\!}S_1$
and $^{3\!}D_1$ states are coupled. The only two free parameters are tuned so
as to reproduce mass and width of the $\phi(1020)$, but come out reasonably
close to previously used values. Among the model's $T$-matrix poles, there
are good candidates for observed resonances, as well other ones that should
exist according to the quark model. Besides the expected resonances as
unitarized confinement states, a dynamical resonance pole is found at
$(2186-i246)$~MeV. The huge width makes its interpretation as the
$\phi(2170)$ somewhat dubious, but further improvements of the model may
change this conclusion.
\end{abstract}


\maketitle

\section{Introduction}
In 2006, the BABAR Collaboration announced \cite{PRD74p091103} the
discovery of a new vector-meson resonance, called $X(2175)$, in the 
initial-state-radiation process $e^+e^-\to K^+K^-\pi\pi\gamma$, observed
in the channel $\phi(1020)f_0(980)$, with the $\phi$ meson decaying to
$K^+K^-$ and the $f_0(980)$ to $\pi^+\pi^-$ or $\pi^0\pi^0$. Two years
later, the BES Collaboration confirmed \cite{PRL100p102003} this resonance,
then denoted $Y(2175)$, in the decay
$J/\psi\to\eta[\to\!\gamma\gamma]\,\phi[\to\!K^+K^-]\,f_0(980)[\to\!\pi^+\pi^-]$.
At present, the new state is included in the PDG listings as
the $\phi(2170)$ \cite{PLB667p1},
though not in the summary table, with average mass $M=(2175\pm15)$~MeV and
width $\Gamma=(61\pm18)$~MeV. However, these resonance parameters are being
strongly challenged by the very recent Belle \cite{PRD80p031101} results on the
$Y(2175)$, alias $\phi(2170)$, observed in the process
$e^+e^-\to\phi\,\pi^+\pi^-$, yielding $M=(2079\pm13^{+79}_{-28})$~MeV and
$\Gamma=(192\pm23^{+25}_{-61})$~MeV.

The observation of this highly excited $\phi$-type resonance with (probably)
modest width, besides the peculiar, seemingly preferential, decay mode
$\phi f_0(980)$, triggered a variety of model explications, most of which
proposing exotic solutions. Let us mention first a strangeonium-hybrid
($s\bar{s}g$) assignment, in the flux-tube as well as the constituent-gluon
model \cite{PLB650p390}, and a perturbative comparison of $\phi(2170)$ decays
in these exotic ansatzes with a standard $2\,{}^{3\!}D_1$ $s\bar{s}$
description from both the flux-tube and the ${}^{3\!}P_0$ model, by the same
authors \cite{PLB657p49}. Other approaches in terms of exotics, with QCD sum
rules, are an $ss\bar{s}\bar{s}$ tetraquark assignment \cite{NPA791p106}, and
an analysis \cite{PRD78p034012} exploring both $ss\bar{s}\bar{s}$ and
$s\bar{s}s\bar{s}$ configurations. In an effective description based on
Resonance Chiral Perturbation Theory \cite{PRD76p074012}, the bulk of the
experimental data is reproduced except for the $\phi(2170)$ peak. This then
led to a 3-body Faddeev calculation \cite{PRD78p074031}, with the pair
interactions taken from the chiral unitary approach. Indeed, a resonance with
parameters reasonably close to those of the $\phi(2170)$ is thus generated, 
though a little bit too narrow. Finally, a review on several puzzling hadron
states \cite{IJMPE17p283} mentions the possibility that the $\phi(2170)$ 
arises from $S$-wave threshold effects.

In the present paper, we shall study the possibility that the $\phi(2170)$
is a normal excited $\phi$ meson, by coupling a complete confinement spectrum
of $s\bar{s}$ states
to a variety of $S$- and $P$-wave two-meson channels, composed of pairs of
ground-state pseudoscalar (P), vector (V), scalar (S), and axial-vector (A)
mesons. The employed formalism is a multichannel generalization of the
Resonance-Spectrum Expansion (RSE) \cite{IJTPGTNO11p179,AOP324p1620}, which 
allows for an arbitrary number of confined and scattering channels
\cite{0905.3308}. 

In Sec.~2 the RSE is very briefly reviewed and the explicit $T$-matrix for
the present coupled-channel system is given. Resonance poles and their
trajectories are shown in Sec.~3, and model cross sections in Sec.~4.
We draw our conclusions in Sec.~5 and discuss possible improvements.

\section{The RSE applied to \boldmath{$\phi$} recurrences}
 
The RSE model has been developed for meson-meson scattering in non-exotic
channels, whereby the intermediate state is described via an infinite tower
of $s$-channel $q\bar{q}$ states. For the spectrum of the latter, in principle
any confinement potential can be employed, but in practical applications, a
harmonic oscillator (HO) with constant frequency has been used, with excellent
results. For more details and further references, see Refs.\
\cite{IJTPGTNO11p179,AOP324p1620,PRL91p012003,PLB641p265,EPJA31p468,PRL97p202001}.

In the present investigation of strangeonium vector mesons, both the \tso\ and
\tdo\ $s\bar{s}$ confinement channels are included, to be compared with recent
work \cite{0905.3302} restricted to the \tso\ component only. We could in
principle also consider deviations from ideal mixing, by coupling the
corresponding two $(u\bar{u}+d\bar{d})/\sqrt{2}$ channels as well, but such
fine corrections will be left for possible future studies. For the meson-meson
channels, we consider the most relevant combinations of ground-state P, V, S,
and A mesons that have nonvanishing coupling to either of the two confinement
channels in accordance with the $^{3\!}P_0$ model and the OZI rule. The
resulting 17 channels are listed, with all their relevant quantum numbers, in
Table~\ref{MM}. For the channels containing an $\eta$ or $\eta'$ meson, we 
assume a pseudoscalar mixing angle of $37.3^\circ$, in the flavor basis,
though our results are not very sensitive to the precise value. Also note that
channels with the same particles but different
relative orbital angular momentum $L$ or total spin $S$ are considered
different. This is only strictly necessary for different $L$, because of the
corresponding wave functions, but is also done when $S$ is different, for the
purpose of clarity. All relative couplings have been computed using the
formalism of Ref.~\cite{ZPC21p291}, based on overlaps of HO wave functions.
They are given in Table~\ref{MM} for the lowest recurrences ($n=0$). As a
matter of fact, we list their squares, which are rational numbers, but given
as rounded floating-point numbers in the table, also for clarity's sake.
For higher $n$ values, the couplings fall off very rapidly. Their $n$
dependence, for the various sets of decay channels, is presented in
Table~\ref{gn}. The threshold values in Table~\ref{MM} are obtained by taking
the meson masses given in the PDG tables or listings \cite{PLB667p1}, with
the exception of the $K_0^*(800)$ (alias $\kappa$), for which we choose the
real part of the pole position from Ref.~\cite{PLB641p265}, as it lies closer
to the world average of $\kappa$ masses. Note that we take sharp thresholds,
even when (broad) resonances are involved. We shall come back to this point
in the conclusions. Finally, we should notice that a number of channels that
also couple to $s\bar{s}$ vector states according to the scheme of
Ref.~\cite{ZPC21p291}, viz.\ $P$-wave channels involving axial-vector mesons
as well as some channels with tensor mesons, have not been included in the 
final calculations presented here. However, their influence has been tested
and turned out to be very modest, due to the corresponding small couplings.

Coming now to the explicit expressions for our model, let us first write
down the effective meson-meson interaction, which consists of an
intermediate-state $s$-channel $q\bar{q}$ propagator between two
quark-antiquark-meson-meson vertex functions for the initial and final state,
reading \cite{AOP324p1620,0905.3308}
\begin{equation}
V_{ij}^{L_i,L_j}(p_i,p'_j;E)=
\lambda^2j^i_{L_i}(p_ia)\mathcal{R}_{ij}(E)j^j_{L_j}(p'_ja) \;,
\label{inter}
\end{equation}
with
\begin{equation}
\mathcal{R}_{ij}(E)=\sum_{l_c=0,2}\sum_{n=0}^{\infty}
\frac{g^i_{(l_c,n)}g^j_{(l_c,n)}}{E-E_n^{(l_c)}}\;,
\label{rse}
\end{equation}
where the RSE propagator contains an infinite tower of $s$-channel bare
$q\bar{q}$ states, corresponding to the spectrum of an, in principle,
arbitrary confining potential. Here, $E_n^{(l_c)}$ is the discrete energy
of the $n$-th recurrence in the $s\bar{s}$ channel with angular momentum
$l_c$, and $g^i_{(l_c,n)}$ is the corresponding coupling to the $i$-th
meson-meson channel. Furthermore, in Eq.~(\ref{inter}), $\lambda$ is an
overall coupling, and $\bes{i}(p_i)$ and
$p_i$ are the $L_i$-th order spherical Bessel function and the
(relativistically defined) off-shell relative momentum in
meson-meson channel $i$, respectively. The spherical Bessel function
originates in our string-breaking picture of OZI-allowed decay,
being just the Fourier transform of a spherical delta function of radius $a$.
Together with the overall coupling constant $\lambda$, the radius $a$ is a
freely adjustable parameter here, though its range of allowed values turns out
to be quite limited in practice. The couplings $g^i_{(l_c,n)}$ in
Eq.~(\ref{rse}) are obtained by multiplying the ones in Table~\ref{MM} by those
in Table~\ref{gn}, for the corresponding channels. Because of the fast decrease
of the latter for increasing $n$, practical convergence of the infinite sum
in Eq.~(\ref{rse}) is achieved by truncating it after 20 terms.

Because of the separable form of the effective meson-meson interaction in
Eq.~(\ref{inter}), the fully off-shell $T$-matrix can be solved in
closed form with straightforward algebra, resulting in the expression
\begin{eqnarray}
\lefteqn{\tmat{i}{j}(p_i,p'_j;E)=-2a\lambda^2\sqrt{\mu_ip_i}\,\bes{i}(p_ia)
\times} \nonumber \\
&&\hspace*{-1pt}\sum_{m=1}^{N}\rse_{im}\left\{[\One-\Omega\,\mathcal{R}]^{-1}
\right\}_{\!mj}\bes{j}(p'_ja)\,\sqrt{\mu_jp'_j} \; ,
\label{tmat}
\end{eqnarray}
with
\begin{equation}
\Omega_{ij}(k_j)=-2ia\lambda^2\mu_jk_j\,\bes{j}(k_ja)\,\han{j}(k_ja)\,
\delta_{ij}\;, 
\label{omega}
\end{equation}
where $\han{j}(k_ja)$ is the spherical Hankel function of the first kind, 
$k_j$ and $\mu_j$ are the on-shell relative momentum and reduced mass in
meson-meson channel $j$, respectively, and the matrix $\mathcal{R}(E)$ is
given by Eq.~(\ref{rse}).
\begin{table}[t]
\centering
\begin{tabular}{|c||c c|c c | c c|c|}
\hline & & & & & & & \\[-3mm]
 & $g^2_{(l_c=0)}$ & $g^2_{(l_c=2)}$ & &  &  &  & Threshold \\[1mm]
Channel & $\times10^{-3}$ & $\times10^{-3}$ & $l_1$ & $l_2$ & $L$ & $S$ &
(MeV) \\ 
\hline & & & & & & & \\[-3mm]
$KK$             & $27.8$  & $9.26$  & $0$&$0$&$1$&$0$ & $987$\\ \hline
$KK^*$           & $111$   & $9.26$  & $0$&$0$&$1$&$1$ & $1388$\\ 
$\eta\phi$       & $40.8$  & $3.40$  & $0$&$0$&$1$&$1$ & $1567$\\ 
$\eta'\phi$      & $70.3$  & $5.86$  & $0$&$0$&$1$&$1$ & $1977$\\ \hline
$K^*K^*$         & $9.26$  & $3.09$  & $0$&$0$&$1$&$0$ & $1788$\\
$K^*K^*$         & $185$   & $0.62$  & $0$&$0$&$1$&$2$ & $1788$\\ \hline
$\phi(1020)f_0(980)$  & $83.3$  & $0$     & $0$&$1$&$0$&$1$ & $1999$\\ 
$K^*K_0^*(800)$  & $83.3$  & $0$     & $0$&$1$&$0$&$1$ & $1639$\\
$\phi(1020)f_0(980)$  & $0$     & $14.7$  & $0$&$1$&$2$&$1$ & $1999$\\ 
$K^*K_0^*(800)$  & $0$     & $14.7$  & $0$&$1$&$2$&$1$ & $1639$\\ \hline
$\eta h_1(1380)$ & $10.2$  & $5.67$  & $0$&$1$&$0$&$1$ & $1928$\\ 
$\eta' h_1(1380)$& $17.6$  & $9.76$  & $0$&$1$&$0$&$1$ & $2338$\\ 
$KK_1(1270)$     & $83.3$  & $20.6$  & $0$&$1$&$0$&$1$ & $1764$\\
$KK_1(1400)$     & $0$     & $2.57$  & $0$&$1$&$0$&$1$ & $1894$\\ \hline
$K^*K_1(1270)$   & $167$   & $10.3$  & $0$&$1$&$0$&$1$ & $2164$\\ 
$K^*K_1(1400)$   & $0$     & $1.29$  & $0$&$1$&$0$&$1$ & $2294$\\
$\phi f_1(1420)$ & $111$   & $3.86$  & $0$&$1$&$0$&$1$ & $2439$\\ \hline
\end{tabular}
\caption{Included two-meson channels, their internal and relative angular
momenta and spins, couplings squared for $n=0$, and thresholds. See
Ref.~\cite{PLB667p1} for properties of listed mesons, except for the
$K_0^*(800)$, discussed in the text.}
\label{MM}
\end{table}
\begin{table}
\centering
\begin{tabular}{|c|c|c|}
\hline & & \\[-3mm]
Channel & $\tilde{g}^2_{(l_c=0,n)}\times4^n$ &
          $\tilde{g}^2_{(l_c=2,n)}\times4^n$ \\ 
\hline & & \\[-3mm]
PP & $(2n+3)/3$ & $n+1$ \\
PV & $(2n+3)/3$ & $n+1$ \\
VV & $(2n+3)/3$ & $n+1$ \\
SV & $(2n-3)^2/9$ & $(n+1)(2n+5)/5$ \\
PA & $(2n-3)^2/9$ & $(n+1)(2n+5)/5$ \\
VA & $(2n-3)^2/9$ & $(n+1)^2$ \\
\hline
\end{tabular}
\caption{Dependence of couplings squared on recurrence $n$.}
\label{gn}
\end{table}
\begin{table}
\centering
\begin{tabular}{|c||c|c|}
\hline & & \\[-3mm]
$n$ & $l_c=0$ & $l_c=2 $\\ [0.5ex]
\hline & & \\[-3mm]
0 & 1301 & 1681\\
1 & 1681 & 2061\\
2 & 2061 & 2441\\
3 & 2441 & 2821\\
\hline
\end{tabular}
\caption{Masses of bare $s\bar{s}$ states in MeV, for HO potential with
$\omega=190$~MeV and $m_s=508$ MeV (see Eq.~(\ref{ho}) and
Ref.~\cite{PRD27p1527}).}
\label{HO}
\end{table}
 
As mentioned above, we assume an HO confinement spectrum with constant
frequency, given by
\begin{eqnarray}
E_n=2m_s+\omega(2n+l_c+1.5) \; ,
\label{ho}
\end{eqnarray}
with $\omega=190$~MeV and $m_s=508$~MeV fixed at values determined long ago
\cite{PRD27p1527}.
Some of the resulting bare HO $s\bar{s}$ states are given in Table~\ref{HO}.

Now that the model has been fully defined, we are in a position to evaluate
the on-shell components of the $T$-matrix defined in
Eqs.~(\ref{tmat},\ref{rse}), for the channels given in
Tables~\ref{MM}--\ref{HO}.

\section{Experimental status of $\phi$ states}

Before adjusting our two free parameters $\lambda$ and $a$ from
Eq.~(\ref{tmat}), let us first have a look at the experimental status of vector
$\phi$ resonances. According to the 2008 PDG listings \cite{PLB667p1}, there
are only 3 observed states, viz. the $\phi(1020)$, $\phi(1680)$, and
$\phi(2170)$, with the latter resonance omitted from the summary table. Their
PDG masses and widths are given in Table~\ref{phis}. Clearly, this is a very
\begin{table}[t]
\centering
\begin{tabular}{|c||c|c|}
\hline & & \\[-3mm]
  & $M$ (MeV) & $\Gamma$ (MeV) \\ 
\hline & & \\[-3mm]
$\phi(1020)$ & 1019.455$\pm$0.020 & 4.26$\pm$0.04\\
$\phi(1680)$ & 1680$\pm$20 & 150$\pm$50\\
$\phi(2170)$ & {\boldmath$2175\pm15$} & {\boldmath $61\pm18$}\\
\hline
\end{tabular}
\caption{Listed $J^{PC}=1^{--}$ $\phi$ resonances, with masses and widths 
\cite{PLB667p1} (values for $\phi(1680)$ are estimates \cite{PLB667p1}).}
\label{phis}
\end{table}
poor status, as several additional states must exist in the energy range
1--2~GeV according to the quark model, and also if we compare with e.g.\
observed $\rho$ resonances \cite{PLB667p1} in the same energy interval.
Moreover, the $\phi(1680)$ can hardly be the first radial excitation of the
$\phi(1020)$, in view of the well established $K^*(1410)$, which is almost
300~MeV lighter, and a typical mass difference of 100--150 MeV between
the strange and nonstrange ($u,d$) constituent quarks
\cite{PRD27p1527,IJMPA13p657}. This conclusion is further supported if indeed
the $\rho(1250)$ is confirmed as the first radial recurrence of the $\rho(770)$
\cite{NPA807p145,PRD27p1527}. So the $\phi(1680)$ is more likely to be the
$1\,{}^{3\!}D_1$ state, with a hitherto undetected $2\,{}^{3\!}S_1$ state
somehwere in the mass range 1.5--1.6~GeV. As a matter of fact, in
Ref.~\cite{PRD57p4334} a vector $\phi$ resonance was reported at roughly
1.5~GeV, though this observation is, surprisingly, included under the
$\phi(1680)$ entry \cite{PLB667p1}. Even more oddly, another
$\phi$-like
state, at $\sim\!1.9$~GeV and reported in the same paper \cite{PRD57p4334},
is {\em also} \/included under the $\phi(1680)$ \cite{PLB667p1}. However,
a resonance at about 1.9~GeV should be a good candidate for the next radial
$s\bar{s}$ recurrence, if we take the observed $\rho$ resonances in
Ref.~\cite{NPA807p145} for granted.

\section{Hunting after poles}
In view of the poor status of excited $\phi$ states, let us adjust our
parameters $\lambda$ and $a$ to the mass and width of the $\phi(1020)$.
Here, we should mention that an additional phenomenological ingredient of
our model is an extra suppression of subthreshold contributions, using a 
form factor, on top of the natural damping due to the spherical Bessel
and Hankel functions in Eq.~(\ref{rse}). Such a procedure is common practice
in multichannel phase-shift analyses. Thus, for closed meson-meson channels
we make the substitution
\begin{equation}
\left(g^i_{(l_c,n)}\right)^2\;\to\; \left(g^i_{(l_c,n)}\right)^2
e^{\alpha k_{i}^{2}}
\;\;\;\;\mbox{for}\;\;\;\;
\Re\mbox{e}\, k_{i}^{2}<0
\; .
\label{damping}
\end{equation}
The parameter $\alpha$ is chosen at exactly the same value as in previous work
\cite{PLB641p265,PRL97p202001}, viz.\ $\alpha=4$~GeV$^{-2}$.

Choosing now $\lambda=4$~GeV$^{-3/2}$ and $a=4$~GeV$^{-1}$, we manage to
reproduce mass and width of the $\phi(1020)$ with remarkable accuracy, namely
$M_\phi=1019.5$ MeV and $\Gamma_\phi=4.4$ MeV. Note that these values of
$\lambda$ and $a$ are of the same order of magnitude as in the work mentioned
before \cite{PLB641p265,PRL97p202001}, which dealt with scalar mesons.

In Table~\ref{poles} we collect all resonance poles encountered on the
respective physical Riemann sheets, which correspond to $\Im\mbox{m}\,k_i>0$
for closed channels and $\Im\mbox{m}\,k_i<0$ for open ones. When the latter
conditions are not fulfilled, we call the corresponding Riemann sheets
unphysical. Moreover, we also show here the pole positions obtained
by taking only the \tso\ $s\bar{s}$ channel and switching off the \tdo, for
fixed $\lambda$ and $a$. Focusing for the moment on those poles that originate
\begin{table}[t]
\centering
\begin{tabular}{|c|cc|cc|c|}
\hline & & & & & \\[-3mm]
&\multicolumn{2}{c|}{${}^{3\!}S_1$ only}&
\multicolumn{2}{c|}{${}^{3\!}S_1+{}^{3\!}D_1$}&\\
\hline & & & & & \\[-3mm]
Pole&$\Re$e & $\Im$m & $\Re$e & $\Im$m & Type of Pole \\ 
\hline & & & & & \\[-3mm]
1&$1027.5$ &$-2.7$& $1019.5$ &$-2.2$& conf., $n=0$, $1\,{}^{3\!}S_1$\\
2&$1537$  &$-13$ & $1516$  &$-23$ & conf., $n=1$, $2\,{}^{3\!}S_1$\\
3&   -   &    -  & $1602$  &$ -6$& conf., $n=0$, $1\,{}^{3\!}D_1$\\
4&$1998$  & $-16$& $1932$  &$-24$ & conf.\, $n=2$, $3\,{}^{3\!}S_1$\\
5&   -   &    -  & $1996$  &$-14$ & conf.\, $n=1$, $\,{}2^{3\!}D_1$\\
6&$2397$  & $-214$& {\boldmath$2186$} & {\boldmath $-246$}&continuum \\
7&$2415$  & $-6$& $2371$  & $-29$  & conf., $n=3$, $4\,{}^{3\!}S_1$\\		
8&-      & -     & $2415$  & $-8$  & conf., $n=2$, $3\,{}^{3\!}D_1$ \\
9&$2501$  & $-236$& $2551$  & $-193$  & continuum \\
\hline
\end{tabular}
\mbox{} \\[1mm]
\caption{Complex-energy poles in MeV, for ${}^{3\!}S_1$ $s\bar{s}$ channel
only, and for both ${}^{3\!}S_1$ and ${}^{3\!}D_1$. See text for further
details.}
\label{poles}
\end{table}
in the states of the confinement spectrum (indicated by ``conf.'' in the
table), we see good candidates for the resonances at $\sim\!1.5$~GeV and
$\sim\!1.9$~GeV reported in Ref.~\cite{PRD57p4334}, and possibly also for the
$\phi(1680)$, though our $1\,{}^{3\!}D_1$ state seems somewhat too light. Note,
however, that under the $\phi(1680)$ entry \cite{PLB667p1} in the PDG listings
there is a relatively recent observation \cite{PLB551p27} with a mass of
$(1623\pm20)$~MeV, which is compatible with our pole at 1602~MeV.
Furthermore, the imaginary parts of the confinement poles are generally too
small, except for the $\phi(1020)$. We shall come back to this point in the
conclusions below. Besides the latter poles, also two so-called {\em
continuum} \/poles are found, often designated as {\em dynamical} \/poles,
the most conspicuous of which is the one at $(2186-i246)$~MeV, as the real part
is very close to the mass of the $\phi(2170)$ as measured by BABAR
\cite{PRD74p091103} and BES \cite{PRL100p102003}. However, in view of the much
too large width, even as compared to the Belle \cite{PRD80p031101} value,
considerable caution is urged. Also this point will be further discussed in the
conclusions.

Some words are in place here about our identification of the \tso\ and \tdo\
confinement poles in Table~\ref{poles}. The point is that, rigorously speaking,
these designations only make sense for pure confinement states and, moreover,
without any \tso\,/\,\tdo\ mixing. Now, in our approach, the very mixing is
provided by the coupling to common decay channels. So for any nonvanishing
value of the overall coupling $\lambda$ there are no longer pure \tso\ and
\tdo\ states, while for the physical value of $\lambda$ the mixing is probably
considerable. Moreover, there is no obvious way to tell which pole of a pair
originating in a degenerate confinement state stems from either \tso\ or \tdo.
Therefore, our identification is partly based on the couplings in
Table~\ref{MM}, which on the whole suggest larger shifts for \tso\ than for
\tdo, partly on a comparison with a perturbative approach employed in
Ref.~\cite{PRD21p772} to find poles for small $\lambda$.

\begin{figure}[t]
\resizebox{!}{225pt}{\includegraphics{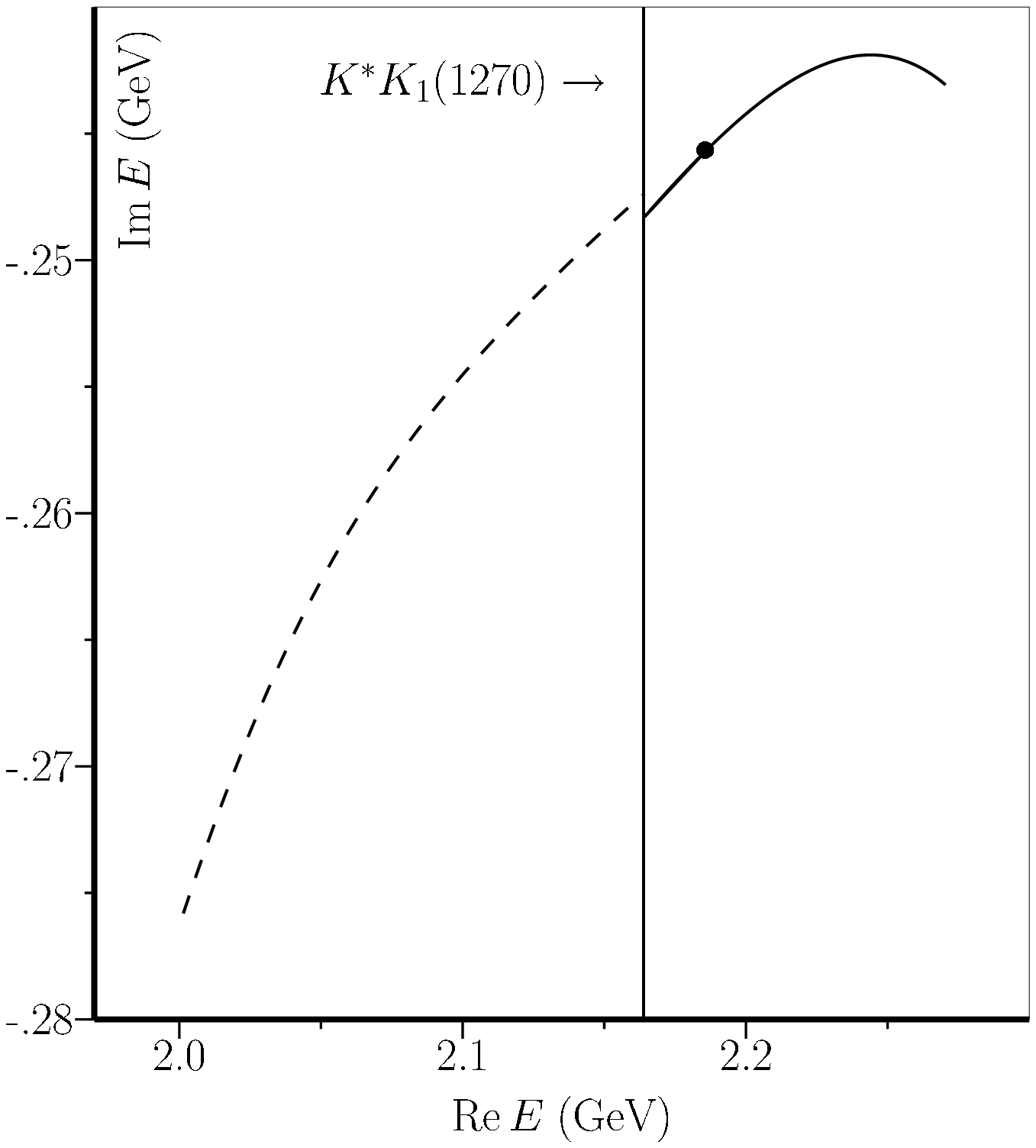}}
\caption{Trajectory of first continuum pole, for $2.26\leq\lambda\leq5.99$
(GeV$^{-3/2}$), from left to right. Bullet represents $\lambda=4$~GeV$^{-3/2}$,
while dashed line indicates unphysical Riemann sheet.}
\label{2170}
\end{figure}

The designation {\em continuum} \/pole becomes clear when plotting a
corresponding trajectory as a function of the overall coupling $\lambda$. In
Fig.~\ref{2170}, the first such pole is shown to have an {\em increasingly} 
\/large imaginary part for {\em decreasing} \/$\lambda$, eventually
disappearing in the continuum for $\lambda\to0$. Note that the small jump
at the important $S$-wave $K^*K_1(1270)$ threshold is due to a minor
threshold discontinuity of the damping function in Eq.~(\ref{damping}) for
complex momenta.

\begin{figure}[h]
\resizebox{!}{225pt}{\includegraphics{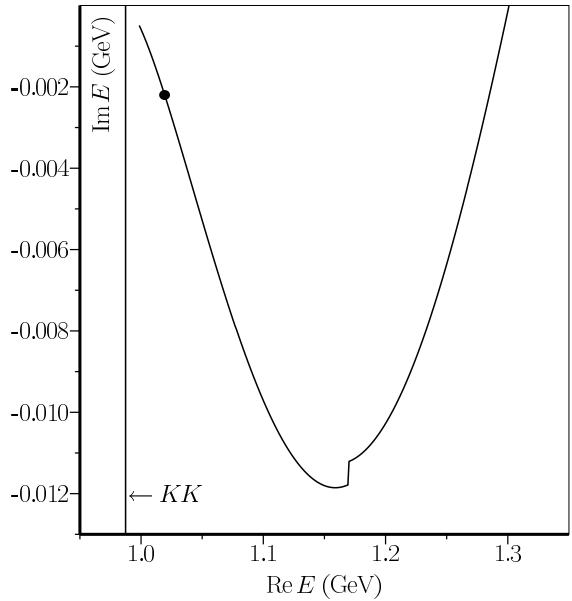}}
\caption{$1\,{}^{3\!}S_1$ confinement pole for $4.31\geq\lambda\geq0$
(GeV$^{-3/2}$). Bullet represents $\lambda=4$~GeV$^{-3/2}$.}
\label{zerothconf} 
\end{figure}

Figure~\ref{zerothconf} shows a similar trajectory, but now for the lowest
confinement pole, which ends up as the $\phi(1020)$ resonance. Notice the
large negative mass shift ($\approx\!280$~MeV), as well as the way the pole
approaches the $K\bar{K}$ threshold, which is typical for $P$-wave decay
channels. Also note that the tiny jump in the trajectory is due to the way 
relativistic reduced mass is defined below threshold, which in the case
of closed channels with highly unequal masses ($KK_1(1270)$ here) requires 
an intervention to prevent the reduced mass from becoming negative.

\begin{figure}[b]
\resizebox{!}{225pt}{\includegraphics{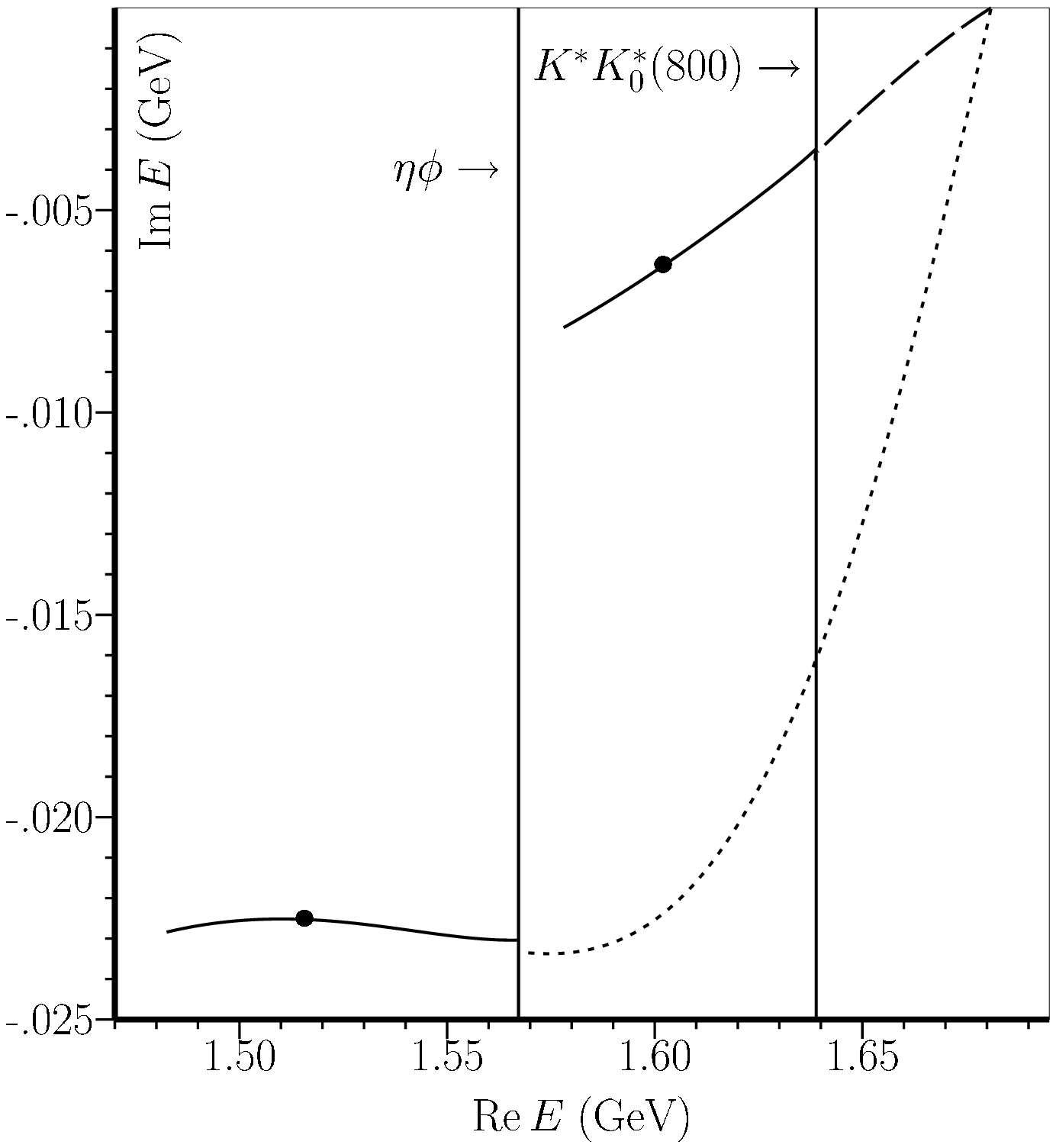}}
\caption{\stso\ (lower) and \ftdo\ (upper) confinement poles for
$5.0\geq\lambda\geq0$ (GeV$^{-3/2}$) and $4.76\geq\lambda\geq0$ (GeV$^{-3/2}$),
respectively. Bullets represent $\lambda=4$~GeV$^{-3/2}$, while dotted and
dashed lines indicate unphysical Riemann sheets.}
\label{firstconf} 
\end{figure}

In Fig.~\ref{firstconf}, we depict the trajectories of the \stso\ and
\ftdo\ confinement poles. Note that the coupling to decay channels lifts
the original degeneracy of the \stso\ and \ftdo\ HO states.

\begin{figure}[t]
\resizebox{!}{225pt}{\includegraphics{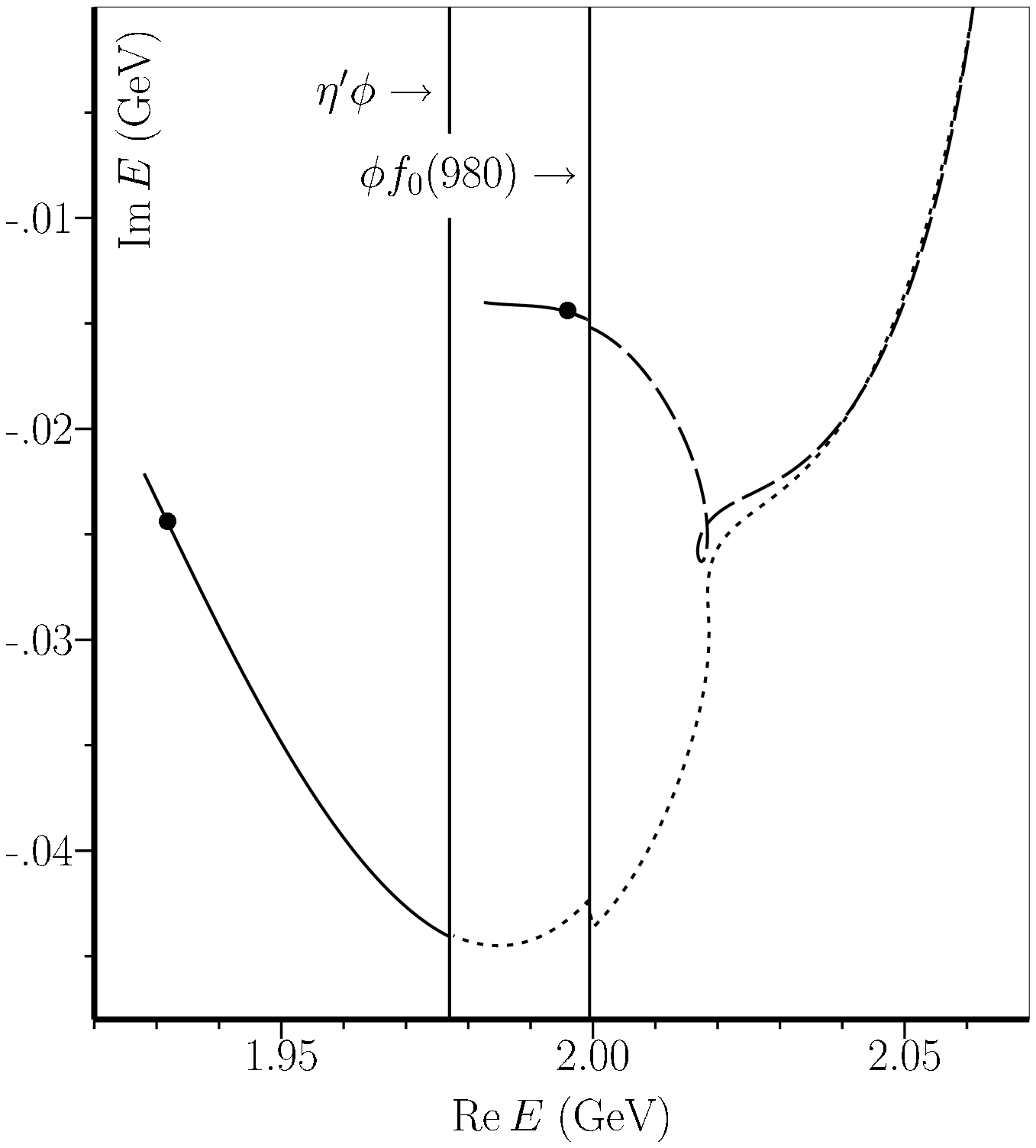}}
\caption{\ttso\ (lower) and \stdo\ (upper) confinement poles for
$4.2\geq\lambda\geq0$ (GeV$^{-3/2}$) and $5.99\geq\lambda\geq0$ (GeV$^{-3/2}$),
respectively. Bullets represent $\lambda=4$~GeV$^{-3/2}$, while dotted and
dashed lines indicate unphysical Riemann sheets.}
\label{secondconf}
\end{figure}

The trajectories of the next pair of confinement poles, i.e., \ttso\ and \stdo,
are drawn in Fig.~\ref{secondconf}. Note the highly nonlinear behavior of the
poles, showing the unreliability of perturbative methods to estimate
coupled-channel effects.

\section{Cross sections}
Now we shall show, as mere illustrations, some of the cross sections related to
the resonance poles found in the preceding section.
\begin{figure}[t]
\resizebox{!}{225pt}{\includegraphics{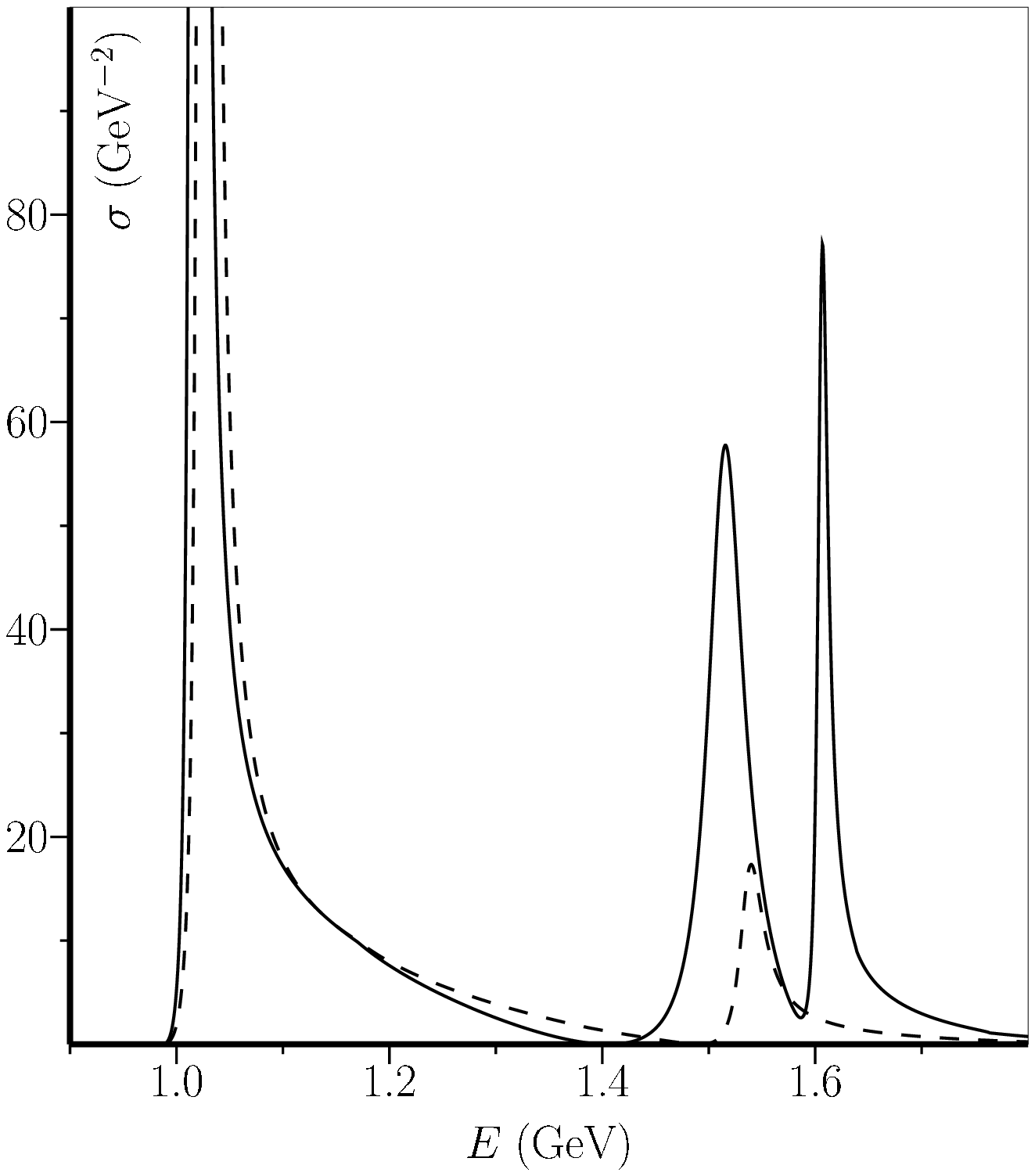}}
\caption{Elastic $P$-wave $KK$ cross section. Full line: both \tso\ and \tdo\
$s\bar{s}$ channels included; dashed line: only \tso.}
\label{KK}
\end{figure}
In Fig.~\ref{KK}, the elastic $P$-wave $KK$ cross section is depicted in
the energy region covering the $\phi(1020)$ as well as the \stso\ and \ftdo\
resonances. We see that including the \tdo\ $s\bar{s}$ channel has the effect
of lowering the \stso\ state, besides the generation of an additional
resonance, of course. This ``repulsion'' between the \tso\ and \tdo\ poles is
also noticed for the \ttso\ and \stdo\ states.

\begin{figure}[t]
\resizebox{!}{225pt}{\includegraphics{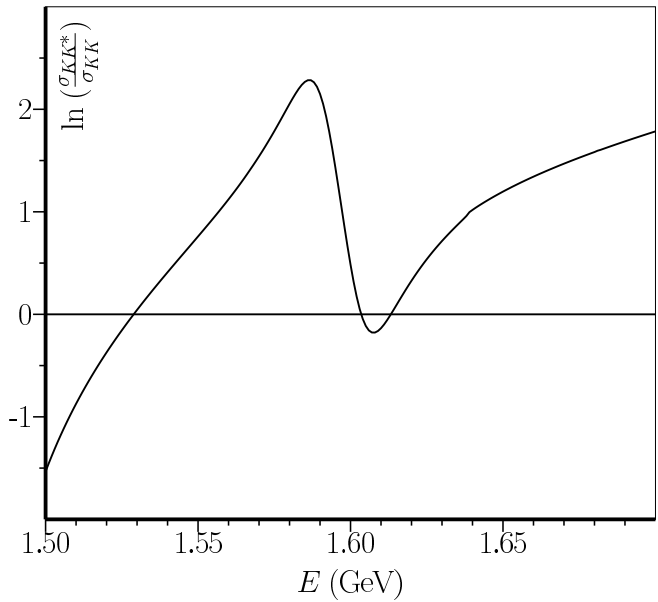}}
\caption{Natural logarithm of the ratio of the elastic $KK^*$ and $KK$
cross sections.}
\label{KKSKK}
\end{figure}

Figure~\ref{KKSKK} shows the relative importance of the $KK$ and $KK^*$
channels in the energy interval 1.5--1.7~GeV, which should be relevant for
the $\phi(1680)$. The plotted quantity is the logarithm of the ratio of
the elastic $KK^*$ and $KK$ cross sections, which shows that the $KK^*$
channel is strongly dominant, except at low energies, because of phase
space, and close to the pole at $\sim\!1.6$~GeV, where the two cross sections
are comparable. Dominance of the $KK^*$ decay mode is reported under the
$\phi(1680)$ PDG entry \cite{PLB667p1}.

\begin{figure}[t]
\resizebox{!}{225pt}{\includegraphics{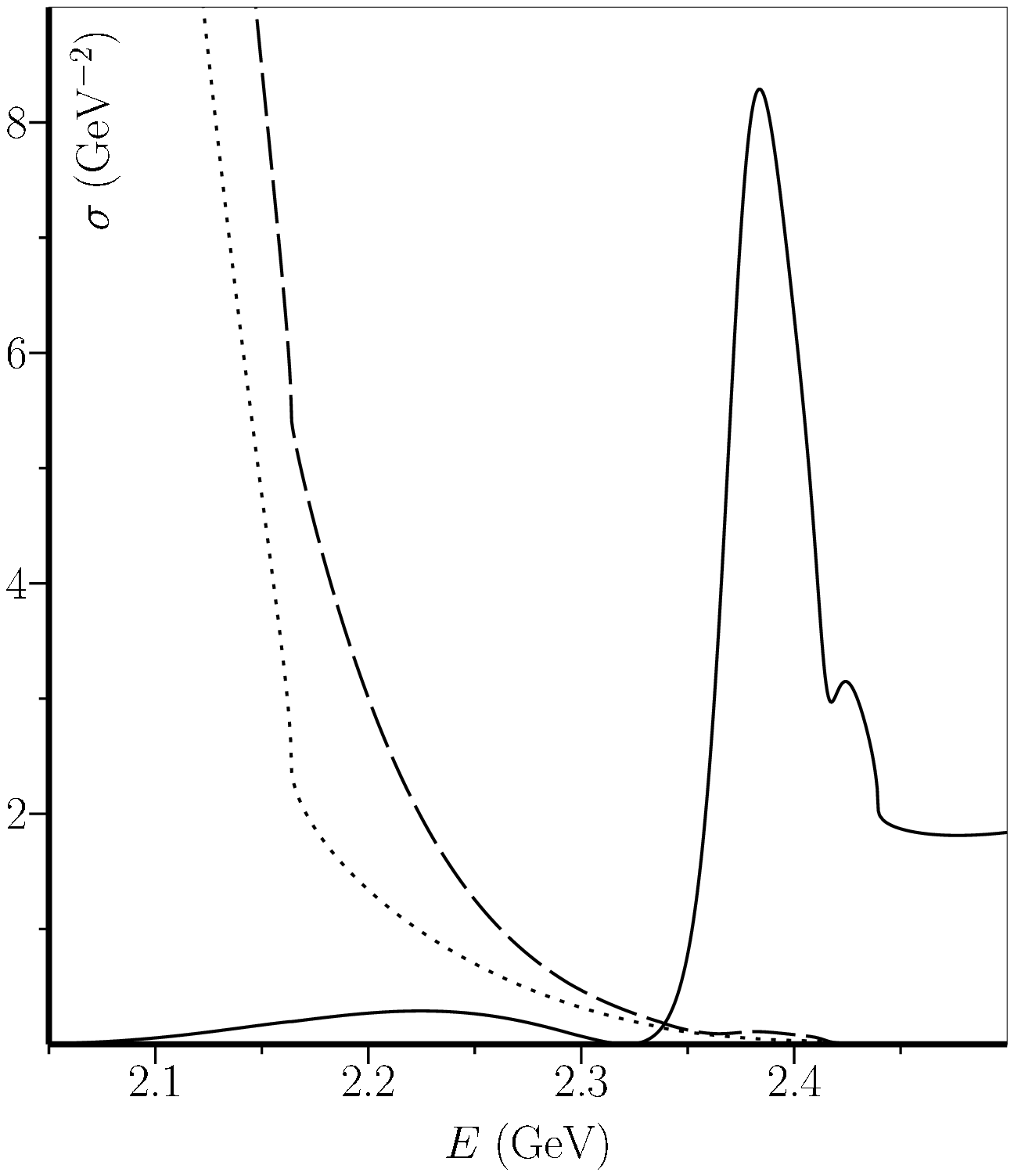}}
\caption{Elastic $D$-wave (solid line) and $S$-wave (dashed line)
$\phi(1020)f_0(980)$ cross section. Dotted line: $S$-wave cross section for
\tso\ channel only.}
\label{f0phi}
\end{figure}

Turning now to the $\phi(2170)$ energy region, we show in Fig.~\ref{f0phi}
the elastic $S$- and $D$-wave
$\phi(1020)f_0(980)$ cross sections. The effect of the continuum pole at
$(2186-i246)$~MeV is noticeable as a small and very broad enhancement in the
$D$-wave cross section. In the $S$-wave case, its effect is completely
overwhelmed by the huge cross section at threshold, partly due to the
\ttso\ pole not far below. Also quite conspicuous are the here predicted
\ftso\ and \ttdo\ resonances (see Table~\ref{poles} for the respective pole
positions). Of course, all these model {\em elastic} \/cross
sections have little direct bearing upon the experimentally observed 
{\em production} \/cross sections. The production process of the
$\phi(2170)$ may be studied with the RSE production formalism 
\cite{AOP323p1215}, but that lies outside the scope of the present
investigation, which focused on the possibility of generating a
$\phi(2170)$ resonance pole through coupled channels.

\begin{figure}[t]
\resizebox{!}{225pt}{\includegraphics{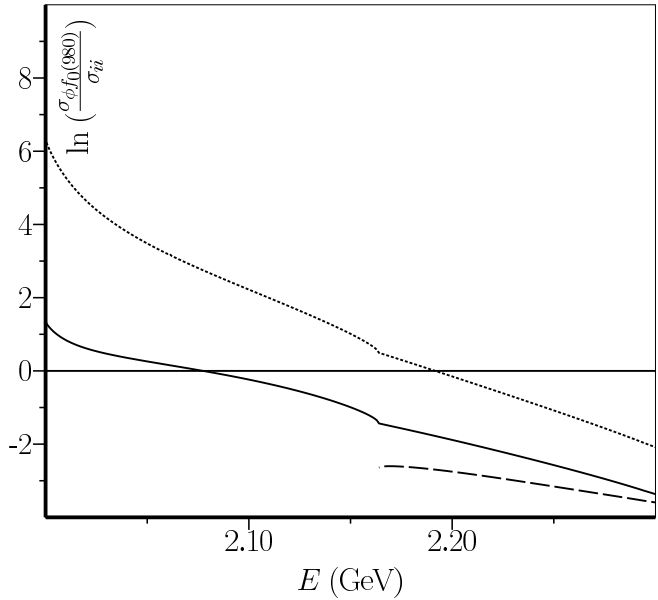}}
\caption{Natural logarithm of the ratios of the elastic $S$-wave
$\phi(1020)f_0(980)$ cross section and the elastic $K^*K^*$ (solid line),
$\phi(1020)\eta'$ (dotted line), and $K^*K_1(1270)$ (dashed line) cross
sections.}
\label{f0phiKK}
\end{figure}

Finally, in Fig.~\ref{f0phiKK} we plot the logarithm of the ratios
of the elastic $S$-wave $\phi(1020)f_0(980)$ cross section and the 
elastic $K^*K^*$, $\phi(1020)\eta'$, and $K^*K_1(1270)$ cross sections,
in the energy interval 2.0--2.3~GeV.
We see that the $S$-wave $\phi(1020)f_0(980)$ cross section dominates
up to about 2.08~GeV, but getting overwhelmed first by the 
($P$-wave) $K^*K^*$ channel, and then even more so by the $S$-wave
$K^*K_1(1270)$ channel, right from its threshold at $\approx\!2.16$~GeV
upwards. Also the $\phi(1020)\eta'$ channel is becoming more important
here. As for the $K^*K^*$ channel, it gives rise to a final state with
two kaons and two pions, i.e., the same as that for which the
$\phi(2170)$ was observed. So the experimental status of the
$\phi(2170)$ might be improved if one succeeded in identifying and
isolating the $K^*[\to K\pi]K^*[\to K\pi]$ decay mode, which should be
quite important.
\section{Summary and conclusions}

In this paper, we have applied the RSE formalism for non-exotic multichannel
meson-meson scattering to calculate the resonance spectrum of excited vector
$\phi$ mesons, and to find out whether this way the $\phi(2170)$ can be
generated. The inclusion of all relevant two-meson channels that couple to
the bare \tso\ and \tdo\ $s\bar{s}$ states should guarantee a reasonable
description.
Thus, several vector $\phi$ resonances are predicted, some of which are good
candidates for observed states, while others may correspond to others,
undetected so far, but quite plausible in view of observed partner states in 
the excited $\rho$ spectrum. Finally, a very broad $\phi$-like resonance pole
of a dynamical origin is found, with real part very close to that of the
$\phi(2170)$, but a much too large imaginary part, so that its interpretation
remains uncertain. On the other hand, the calculated resonances originating in
the confinement spectrum are generally too narrow.

These considerations bring us to the main problem of our description,
namely the inclusion of sharp thresholds only. The point is that many
of the channels in Table~\ref{MM} involve highly unstable particles, several
of which are broad to very broad resonances themselves. Treating the
corresponding thresholds as sharp is clearly an approximation. In particular,
the $f_0(980)$ meson included in the $\phi(1020)f_0(980)$ channels is a
very pronounced resonance in the coupled $\pi\pi$-$KK$ system. This feature
is crucial in the three-body calculation of the $\phi(2170)$ in
Ref.~\cite{PRD78p074031}, which indeed produces a clear resonance signal
at almost the right energy, and even with a somewhat too {\em small} \/width.
We believe that in our approach, too, a narrower $\phi(2170)$
might be generated, if we could account for the physical width of the
$f_0(980)$ meson, and also for the widths of the $K^*$ and
$K_1(1270)$ resonances in the here included $K^*K_1(1270)$ channel. The
reason is that the widths effectively cause these channels to act already
below their central thresholds, which will strongly influence poles just
underneath. Especially the width of very strongly coupling $K^*K_1(1270)$
channel, whose threshold lies only some 25~MeV below the real part of the
continuum pole at $(2186-i\times246)$~MeV, will surely have a very significant
effect on this pole's trajectory. Because of the typical behavior of continuum
poles, with decreasing width for increasing coupling, we expect that the width
of our $\phi(2170)$ candidate may thus be reduced.  Conversely, including the
widths of final-state resonances will probably {\em increase} \/the
widths of the now too narrow excited $\phi$ resonances stemming from the
confinement spectrum.

Of course, to account for the nonvanishing widths of mesons in the
coupled channels is a very difficult problem, since the simple 
substitution of the here used real masses by the true complex masses
will destroy the manifest unitarity of the $S$-matrix. Work is in progress
to redefine the $S$-matrix for such cases, so as to enforce its unitarity
by construction. \\[5pt]

\begin{acknowledgments}
We are indebted to Dr.~Kanchan Khemchandani for very useful discussions on
the $\phi(2170)$ resonance, and to Prof.~David Bugg for several helpful
comments on the first version of the present paper. This work was supported by
the \emph{Funda\c{c}\~{a}o para a Ci\^{e}ncia e a Tecnologia}
\/of the \emph{Minist\'{e}rio da Ci\^{e}ncia, Tecnologia e Ensino Superior}
\/of Portugal, under contract no.\ CERN/FP/\-83502/\-2008.
\end{acknowledgments}

\end{document}